\documentclass[preprint,showpacs,preprintnumbers,amsmath,amssymb]{revtex4}

\usepackage{graphicx}
\usepackage{bm}% bold math

\begin{document}

\title{Stochastic phase reduction for a general class of noisy limit cycle oscillators}

\author{Jun-nosuke Teramae}
\affiliation{Brain Science Institute, RIKEN, Wako 351-0198, Japan}
%\email{teramae@riken.jp}

\author{Hiroya Nakao}
\affiliation{Department of Physics, Kyoto University, Kyoto 606-8502, Japan}

\author{G. Bard Ermentrout}
\affiliation{Department of Mathematics, University of Pittsburgh,
  Pittsburgh, Pennsylvania 15260, USA}

\begin{abstract}
 We formulate a phase reduction method for a general class of noisy limit
 cycle oscillators and find that the phase equation is parameterized by the
 ratio between time scales of the noise and amplitude-relaxation time of the limit
 cycle. The equation naturally includes previously proposed and mutually
 exclusive phase equations as special cases. The validity of the theory is
 numerically confirmed. Using the method, we reveal how noise and its
 correlation time affect limit cycle oscillations.
\end{abstract}



\maketitle

Self-sustained oscillations are widely observed in physical, chemical
and biological systems \cite{kuramoto84, winfree01, pikovsky01}. The
oscillations are often described as limit cycle oscillators. Since limit
cycle oscillators show rich and varied properties, they have been
extensively studied as a central issue of nonlinear science. Timing of
limit cycle oscillation can be described by a single phase variable. The
phase reduction method is a powerful analytical tool to approximate
high-dimensional limit-cycle dynamics as a closed equation for only the
single phase variable \cite{kuramoto84}. Based on the phase description,
studies have revealed fascinating properties of limit-cycle oscillators
like response properties and their collective dynamics
\cite{ermentrout96, acebron05, kawamura08}.

While the theory of phase reduction has been developed mainly for
deterministic limit cycle oscillators, oscillators in the real world are
often exposed to noise. Sources of the noise can be internal
fluctuations, background noise and also input signals which have
noise-like statistics \cite{ermentrout07}. Since noisy limit cycle
oscillators also show various nontrivial properties, there have been
many recent studies of them
\cite{teramae, pikovsky, nakao, galan07, ermentrout08, teramae08, lin08,
yoshimura08}.
While the phase-reduction method is among the most useful ways to
study the effects of noise on oscillators, two mutually exclusive
phase equations have been proposed for a limit cycle oscillator driven
by white Gaussian noise. The first one is formally the same as the phase
equation obtained from deterministic oscillators and is in a sense a
limiting case of colored noise
\cite{teramae, pikovsky, nakao, galan07, ermentrout08, teramae08}
while the second one has an additional term being proportional to square
of noise strength and is the technically correct phase equation for
white noise \cite{yoshimura08}.

Their relationship and which of them is more appropriate description of 
noisy physical oscillators have not been addressed in the literature.
Rather, it was recently pointed out that both of them fail to describe noisy
oscillations in some cases \cite{nakao09}. These facts must imply
existence of a more appropriate phase equation, which will be a starting
point for future research of noisy oscillations. 
In this letter, we solve these problems
by formulating the stochastic phase reduction with careful
consideration of relationship between correlation time of the noise and
relaxation time of the amplitude of the limit cycle.

Noise in the real world has small but finite correlation time
\cite{kampen92}. When the correlation time is much smaller than
characteristic time scales of the noise-driven system, we can use the
white noise description by taking the limit where the correlation time
goes to zero. For limit cycle oscillators, this condition might seem to
mean that the correlation time is much smaller than the period of
oscillation. However, limit cycle oscillators always have other
significant time scales, i.e., the rate of attraction of perturbations
to the limit cycle. These rates characterize
stability of the limit cycle against amplitude perturbation. When
the limit cycle is very stable to perturbations, the decay time
constant could be as small as the short correlation time of
the noise. Since interplay of small time constants can play a crucial
role in stochastic dynamical systems, we should carefully consider their
relationship when we take the white noise limit for noisy limit cycle
oscillators. We employ an Ornstein-Uhlenbeck process which
explicitly has a finite time correlation and then take the white noise
limit of the process while at the same time keeping track of the time
constant for attraction to the limit cycle. 

Let us consider a smooth limit cycle oscillator driven by the
Ornstein-Uhlenbeck process with the time constant $\tau_\eta$,
%% eq.x
\begin{align}
 \begin{split}
  \dot{\bm{X}} &= \bm{F}(\bm{X}) + \sigma \bm{G}(\bm{X}) \eta(t) \\
  \tau_\eta \dot{\eta} &= - \eta + \xi(t),
 \end{split}
 \label{x}
\end{align}
where $\bm{X}(t) \in \bm{R}^N$ is the state of the oscillator at time $t$,
$\bm{F}(\bm{X})$ is its intrinsic dynamics, $\bm{G}(\bm{X})$ is a vector
function, $\xi(t)$ is the zero mean white Gaussian noise of unit
intensity, $\langle \xi(t) \rangle =0$ and $\langle\xi(t)\xi(s)\rangle
=\delta(t-s)$, and then $\eta(t)$ represents the zero mean
Ornstein-Uhlenbeck process with correlation time $\tau_\eta$,
$\langle\eta(t)\eta(s)\rangle =\exp( - |t-s| / \tau_\eta)/(2\tau_\eta)$.
As we take the limit $\tau_\eta \to 0$, $\eta(t)$ approaches the white
Gaussian process of unit strength. $\sigma $ represents noise
strength. $\bm{F}(\bm{X})$
has a stable limit cycle solution $\bm{X}_0 (t)$ satisfying
$\dot{\bm{X}_0}=\bm{F}(\bm{X}_0)$ with period $T$, 
$\bm{X}_0 (t+T)=\bm{X}_0 (t)$. The phase variable $\phi$ is defined
around the limit cycle solution and increases by $T$ for every cycle of
$\bm{X}(t)$ along the limit cycle. Thus, intrinsic angular velocity of
the phase is equal to one. We introduce the other $N-1$ dimensional coordinates
$\bm{\rho} =(\rho_1,\rho_2,\dots )$ to describe the $N$ dimensional dynamics of
$\bm{X}$ using the coordinate $(\phi,\bm{\rho})$
\cite{yoshimura08}. Without loss of generality, we can shift the origin
of $\bm{\rho}$ to $\bm{\rho}=0$ on the limit cycle solution. For simplicity of
the analysis, we assume that $N=2$. Generalization of results to any
values of $N$ is straightforward. We now introduce new variable
$y(t)=\eta(t) \sqrt{\tau_\eta}$. Unlike $\eta$, 
$y$ has the steady distribution, $P_0(y)=\exp(-y^2)/\sqrt{\pi}$, which is 
independent of the correlation time $\tau_\eta$. Variable translations from $\bm{X}$ to
$(\phi,\rho)$ and from $\eta$ to $y$ gives
%% eq.phi
\begin{align}
 \begin{split}
  \dot{\phi} &= 1 + \sigma h(\phi,\rho)\frac{y}{\sqrt{\tau_\eta}}\\
  \dot{\rho} &= \frac{1}{\tau_\rho(\phi)} f(\phi,\rho) + \sigma g(\phi,\rho)\frac{y}{\sqrt{\tau_\eta}}\\
  \dot{y} &= - \frac{y}{\tau_\eta} + \frac{\xi(t)}{\sqrt{\tau_\eta}}.
 \end{split}
 \label{phi}
\end{align}
The functions $h$, $f$ and $g$ are defined as
$h(\phi,\rho)=\nabla_{\bm{X}}\phi\cdot\bm{G}(\bm{X})|_{\bm{X}=\bm{X}(\phi,\rho)}$,
$f(\phi,\rho)/\tau_\rho=\nabla_{\bm{X}}\rho\cdot\bm{F}(\bm{X})|_{\bm{X}=\bm{X}(\phi,\rho)}$
and
$g(\phi,\rho)=\nabla_{\bm{X}}\rho\cdot\bm{G}(\bm{X})|_{\bm{X}=\bm{X}(\phi,\rho)}$
\cite{yoshimura08}.
Since the limit cycle at $\rho=0$ is stable, we explicitly introduced amplitude-relaxation
time of the limit-cycle as $\tau_\rho$, which generally depends on $\phi$
and assumed that $f(\phi,0)=0$ and $\partial f(\phi,0)/\partial\rho=-1$.
The value of $\tau_\rho$ can be
very small if the limit cycle is stiff against amplitude perturbations.

To eliminate the amplitude variable $\rho$ and perform the phase reduction,
we assume that the limit cycle is sufficiently stable and take the limit
$\tau_\rho\to 0$. Simultaneously, we have to take the white noise limit
$\tau_\eta\to 0$. To consider these two limits at the same time, we take
the both limits $\tau_\rho\to 0$ and $\tau_\eta\to 0$ simultaneously
keeping the ratio $k=\tau_\eta/\tau_\rho$ constant. Introducing a small
parameter $\epsilon=\sqrt{\tau_\eta}$, we translate the variable $\rho$ to
$r=\rho/\epsilon$, which remains $O(1)$ as $\epsilon\to 0$.
Expanding $h$, $f$ and $g$ as 
$h(\phi,\epsilon r)=h_0(\phi)+h_1(\phi)\epsilon r+h_2(\phi)\epsilon^2 r^2+\dots$,
$f(\phi,\epsilon r)=-\epsilon r+f_2(\phi)\epsilon^2 r^2+f_3(\phi)\epsilon^3 r^3+\dots$
and 
$g(\phi,\epsilon r)=g_0(\phi)+g_1(\phi)\epsilon r+g_2(\phi)\epsilon^2 r^2+\dots$, 
we obtain the Fokker-Planck equation \cite{horsthemke84, gardiner86} for the 
distribution function $Q(\phi,r,y,t)$ from the stochastic differential equation Eq.(\ref{phi})
as
%% eq.fp
\begin{align}
 \epsilon^2 \frac{\partial Q}{\partial t}
 = (L_0 -\epsilon L_1 -\epsilon^2 L_2)Q+O(\epsilon^3),
 \label{fp}
\end{align}
where linear operators are defined as 
$L_0 Q = (yQ)_y + Q_{yy}/2 + k(rQ)_r - \sigma y g_0 Q_r$,
$L_1 Q = \sigma y[ g_1(rQ)_r + (h_0 Q)_\phi ] + k f_2 (r^2 Q)_r$
and
$L_2 Q = \sigma y[ g_2(r^2 Q)_r + r (h_1 Q)_\phi ] + Q_\phi + k f_3 (r^3 Q)_r$.
Subscript $x$ means partial derivative with respect to the variable $x$.
We assume that $Q$ vanishes rapidly as $y\to\pm\infty$ or $r\to\pm\infty$.
Expanding $Q$ in a perturbation series in $\epsilon$, 
$Q = Q_0 + \epsilon Q_1 + \epsilon^2 Q_2 + \dots$, and equating coefficients 
of equal power of $\epsilon$ in Eq.~(\ref{fp}), we obtain
%% eq.eps
\begin{align}
 \epsilon^0: L_0 Q_0 &= 0 \label{e0}\\
 \epsilon^1: L_0 Q_1 &= L_1 Q_0 \label{e1}\\
 \epsilon^2: L_0 Q_2 &= \frac{\partial}{\partial t}Q_0 + L_2 Q_0 + L_1 Q_1.
 \label{e2}
\end{align}
The lowest order equation, Eq.~(\ref{e0}), has a solution,
$Q_0=P(\phi,t)W(\phi,r,y)$, 
where 
$W(\phi,r,y)=\sqrt{k}(1+k)/(\sigma g_0 \pi) \exp(-y^2-k(y-(1+k)r/(\sigma g_0))^2)$
is the steady Gaussian distribution function of $r$ and $y$ with frozen $\phi$ and 
$g(\phi,r)=g_0(\phi)$. $P(\phi,t)$ is the distribution
function of the $\phi$. Our primary goal is to find the
evolution equation for $P$, which is nothing but the reduced
Fokker-Planck equation for the phase variable $\phi$
\cite{horsthemke84, gardiner86}.

Since the linear operator $L_0$ has the zero eigenvalue, Eq.~(\ref{e1})
and (\ref{e2}) have to fulfill a solvability condition known as the Fredholm alternative.
That is, $L_0 U = b$ has a solution if and only if, $b$ is orthogonal to 
the nullspace of the adjoint of $L_0.$  This nullspace is simply the 
constant function 1. Thus we can solve $L_0 U = b$ when the integral of 
$b$ over $(r,y)$ vanishes.
To obtain this condition, we integrate both sides of these equations
with respect to both $r$ and $y$ from $-\infty$ to $\infty$. We will see that the 
condition for Eq.~(\ref{e2}) is nothing but the desired Fokker-Planck equation for $\phi$. 
Equation (\ref{e1}) is solvable since integration over $(r,y)$ is zero. To see 
why, note that integration of the term $(r Q_0)_r$ with respect to $r$ 
vanishes since $r Q_0(r,y)$ vanishes as $|r|\to\infty.$  Integration of 
$yQ_0(y,r)$ first with repect to $r$ yields an odd function of $y$ which 
is absolutely integrable and thus its integral over $y$ vanishes. We do 
not need the full expression for $Q_1$ at this point, so defer its 
calculation to the next step.
Integration of Eq.~(\ref{e2}) gives
%% eq.sc
\begin{align}
 0= P_t 
   + \sigma \left[h_0\int_{-\infty}^{\infty}\int_{-\infty}^{\infty} (yQ_1) drdy 
                        + \frac{\sigma g_0}{2(1+k)} h_1 P\right]_\phi
   + P_\phi,
   \label{sc}
\end{align}
where we used the rapidly vanishing assumption of $Q$. 
The coefficient of the 3rd term comes from the relationship 
$\int_{-\infty}^{\infty}\int_{-\infty}^{\infty} (yrW) drdy=\sigma g_0/\left(2\left(1+k\right)\right)$,
which is the correlation between $y$ and $r$ for fixed $\phi$.
To evaluate $\int_{-\infty}^{\infty}\int_{-\infty}^{\infty} (yQ_1) drdy$
of the 2nd term, we integrate Eq.~(\ref{e1}) with respect to
$r$ from $-\infty$ to $\infty$ and obtain
%% eq.deq
\begin{align}
 \left(y\int_{-\infty}^{\infty}Q_1 dr\right)_y
 +\frac{1}{2}\left(\int_{-\infty}^{\infty}Q_1 dr\right)_{yy}
 =\frac{\sigma\left(h_0 P\right)_\phi}{\sqrt{\pi}}ye^{-y^2}.
 \label{deq}
\end{align}
Since Eq.~(\ref{deq}) is a differential equation for $\int_{-\infty}^{\infty}Q_1 dr$ 
with respect to $y$, we obtain
$\int_{-\infty}^{\infty}Q_1 dr=-\sigma\left(h_0 P\right)_\phi ye^{-y^2}/\sqrt{\pi}$
by solving this equation. Then we find that
%% eq.intq1
\begin{align}
 \int_{-\infty}^{\infty}\int_{-\infty}^{\infty} (yQ_1) drdy=-\frac{\sigma}{2}(h_0 P)_\phi.
 \label{intq1}
\end{align}
Substituting Eq.~(\ref{intq1}) into Eq.~(\ref{sc}) gives the
partial differential equation for P as,
%% eq.fpp
\begin{align}
 0 = (P_t + P_\phi)
   - \frac{\sigma^2}{2}
     \left[( h_0 ( h_0 P)_\phi )_\phi 
           - \frac{1}{1+k}( h_1 g_0 P )_\phi
     \right],
 \label{fpp}
\end{align}
which is just the Fokker-Planck equation for the phase variable. Finally,
we obtain the phase equation as the Ito stochastic differential
equation equivalent to the Fokker-Planck equation as
%% eq.ito
\begin{align}
 \dot{\phi}= 1 + \frac{\sigma^2}{2} Z_\phi(\phi)Z(\phi)
               + \frac{1}{1+k(\phi)} \sigma^2 Y(\phi)
               + \sigma Z(\phi)\xi(t),
 \label{ito}
\end{align}
where we introduce $Z(\phi)=h_0(\phi)=h(\phi,0)$ and
$Y(\phi)=h_1(\phi)g_0(\phi)/2=h_r(\phi,0)g(\phi,0)/2$.
This is also equivalent to the stochastic differential equation
%% eq.str
\begin{align}
 \dot{\phi}= 1 + \frac{1}{1+k(\phi)} \sigma^2 Y(\phi)
               + \sigma Z(\phi)\xi(t),
 \label{str}
\end{align}
in the Stratonovich interpretation.

We now examine the consequence of the above result. The obtained phase
equation is explicitly parameterized by the ratio between time
constants, $k=\tau_\eta/\tau_\rho$. When the correlation time of the noise
is much smaller than the decay time constant, we can assume $k=0$ and
Eq.~(\ref{str}) is reduced to
$\dot{\phi}=1+\sigma^2 Y(\phi)+\sigma Z(\phi)\xi(t)$,
which is just the phase equation proposed by Yoshimura and Arai
\cite{yoshimura08}. This implies
that when noise is white Gaussian noise in the strict sense, the 2nd
term $Y(\phi)$ must be included in the phase equation. On the
other hand, when the amplitude of the limit cycle decays much faster than
the correlation time of the noise, or the limit-cycle is sufficiently
stable against amplitude perturbations, we can assume that
$k=\infty$ and the 2nd term vanishes. Thus Eq.~(\ref{str}) is reduced to
$\dot{\phi}=1+\sigma Z(\phi)\xi(t)$, which is the same to the equation used in
\cite{teramae, pikovsky, nakao, galan07, ermentrout08, teramae08}.
The latter equation is directly obtained if we apply the
standard phase reduction method to
$\dot{\bm{X}}=\bm{F}(\bm{X})+\sigma\bm{G}(\bm{X})\xi(t)$ without concern for
stochastic nature of the perturbation \cite{kuramoto84}. Thus, the above
result ensures
that we can formally use the standard phase reduction in these
cases. While Eq.~(\ref{str}) agrees with previously proposed
equations at opposite limits of the parameter $k$, it deviates
from both of them in the middle range of $k$. Therefore, we can conclude
that in order to properly describe stochastic phase dynamics for a
general value of $k$, we must consider the coefficient of the 2nd term
correctly as $1/(1+k)$ in the phase equation.

To see the effect of the weight $1/(1+k)$, we will calculate the steady
distribution function for the phase. Requiring the steady condition
$P_t=0$ to Eq.~(\ref{fpp}), we obtain the steady distribution as:
%% eq.p0
\begin{align}
 P_0(\phi) = \frac{1}{T}
             \left(1 + \sigma^2
                  \left[\frac{Z_\phi(\phi)Z(\phi)}{2}
                        -\frac{Y(\phi)}{1+k(\phi)}+\Omega_0
                  \right]
             \right)
             + O(\sigma^4),
 \label{p0}
\end{align}
where we used power series expansion of the distribution in terms of
$\sigma^2$. $\Omega_0$ is defined as
$\Omega_0=T^{-1}\int_0^T Y(\phi)/(1+k(\phi))d\phi$.
As we increase noise strength $\sigma$ from zero, the phase distribution
starts to deviate from $1/T$ of non-perturbed oscillators. While
magnitude of the deviaton is a function of $\sigma$, actual shape of
this depends on the ratio $k(\phi)$.

Using the steady distribution, we can calculate the mean frequency of
the noisy oscillator defined as $\Omega=\lim_{t\to\infty}t^{-1}\int_0^t
\dot{\phi}(t)dt$. Replacing the long term average with the ensemble
average,
i.e.
$\Omega=\int_0^{T}\dot{\phi}P_0(\phi)d\phi$, and substituting the
Ito equation Eq.~(\ref{ito}) into $\dot{\phi}$, we have
%% eq.omg
\begin{align}
 \Omega = 1 + \sigma^2 \Omega_0
            + O(\sigma^2),
 \label{omg}
\end{align}
where we used the fact that $\phi(t)$ is independent from $\xi(t)$ in
the Ito equation. As pointed out in the previous study
\cite{yoshimura08}, the mean frequency
depends on the noise strength. In addition to the strength, our result
reveals that the frequency also depends on $\tau_\eta$ and $\tau_\rho$
through the ratio $k$. As we change these values, the mean frequency
will increases or decreases depending on the sign of $\Omega_0$.

In order to validate the above analysis, we numerically examine
stochastic phase dynamics and calculate $P_0$ and $\Omega$ directly from
the stochastic differential equation (\ref{x}). As a simple example, we
use the Stuart-Landau (SL) oscillator, $\bm{X}=(x,y)$,
$\bm{F}(\bm{X})=(\Re(Z(W)),\Im(Z(W)))$, where
$W=x+iy$ and $Z(W)=(\lambda(1+ic)+i\omega)W-\lambda(1+ic)\left|W\right|^2 W$, 
which is rescaled such that
amplitude relaxation time will explicitly appear. We define phase and
amplitude coordinates $(\phi,r)$ as
$\phi=(\arctan(y/x)-c\log(x^2+y^2)/2)/\omega$ and
$r=\sqrt{x^2+y^2}-1$. The limit cycle solution $x^2+y^2=1$ is given as
$r=0$ in the coordinate. The decay time constant to the limit cycle
solution is $\tau_\rho=1/(2\lambda)$. Figure 1 shows steady state
distributions of the phase for various values of time constants
$\tau_\eta$ and $\tau_\rho$. As expected, the distribution changes as a
function of time constants. Distributions, however, are the same as
far as the ratio between them is the same.
Numerical results are well fitted by the analytical
result Eq.~(\ref{p0}). Figure 2 shows the mean frequency $\Omega$ as a
function of $\tau_\eta$ and $\tau_\rho$. As indicated by the above
analysis, $\Omega$ increases as a function of $\tau_\eta$ and decreases
as a function of $\tau_\rho$. Theoretical predictions, Eq.~(\ref{omg}),
agree fairly well with the numerical results.

The above results clearly indicate that, when we eliminate fast
variables in stochastic dynamical systems, characteristic time scales of
the fast variables should be seriously considered even though variables
themselves are eventually eliminated. In particular, white Gaussian noise is
actually an idealization of physical processes with small but finite
time correlation. Interactions between small time scales can give
crucial effects to stochastic dynamics. Thus similar situations may also
arise even when we use reduction methods other than the phase reduction
to stochastic phenomena \cite{arnold98}. Actually a similar situation
arises in the analysis of classical Brownian motion with inertia
\cite{kupferman04}. The above results also tell us that dynamical
systems driven by the white-Gaussian noise are derived through reduction
methods not only from literally white-noise-driven systems but also from
systems driven by realistic noise with finite time correlations. The
non-agreement between previously proposed phase equations is due to this
ambiguity. Our results ensure that we can choose the most suitable
reduced equation as far as we explicitly indicate time scales of the
noise and dynamical systems.

In summary, we have formulated stochastic phase reduction for a general
class of smooth limit cycle oscillators. The derived stochastic phase equation is
parameterized by the ratio between the correlation time of the noise and
the decay time of amplitude perturbations. Whereas previously proposed
phase equations are realized only at opposite limits of the ratio, the
obtained phase equation is valid in the whole range of values of the
ratio. We have calculated steady phase distributions and the mean
frequency of the noisy oscillator and reveal their dependence on the
time scales. The results suggest significance of fast time scales in
reduction methods of stochastic phenomena.

JT was supported by Kakenhi (B) 20700304. GBE was supported by a grant 
form the National Science Foundation. We would like to thank an anonymous 
reviewer for fixing flaws in our original calculations.

%%%%%%%%%%%%%%%%%%%%%%%%%%%%%%%%%%%%%%%%%%%%%%%%%%%%%%%%%%%%%%%%%%%%%%
\begin{figure}
 % figure1
 \includegraphics{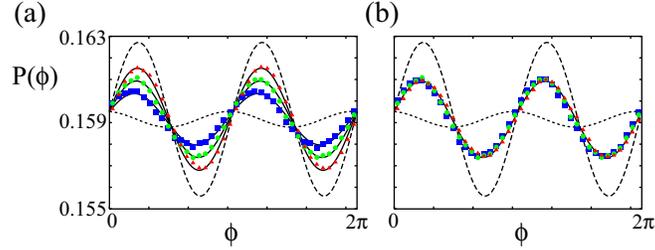}
 \caption{
 Steady distribution function of Stuart-Landau oscillators driven by
 Ornstein-Uhlenbeck processes when $\bm{G}=(1,0)$, $\sigma=0.3$,
 $\Omega=1$ and $c=0.1$. Symbols are numerical results and solid lines are
 theoretical predictions, Eq. (\ref{p0}). Dotted and dashed lines
 are Eq. (\ref{p0}) with $k=0$ and $k=\infty$ respectively. (a)
 $(\tau_\eta,\tau_\rho)=(0.2, 0.1)$ (triangles), (0.1, 0.1) (circles)
 and (0.1, 0.2) (squares). (b) $(\tau_\eta, \tau_\rho)=(0.2, 0.2)$
 (triangles), (0.1, 0.1) (circles) and (0.05, 0.05) (squares).
 }
 \label{figure:1}
\end{figure}
%%%%%%%%%%%%%%%%%%%%%%%%%%%%%%%%%%%%%%%%%%%%%%%%%%%%%%%%%%%%%%%%%%%%%%
%%%%%%%%%%%%%%%%%%%%%%%%%%%%%%%%%%%%%%%%%%%%%%%%%%%%%%%%%%%%%%%%%%%%%%
\begin{figure}
 % figure2
 \includegraphics{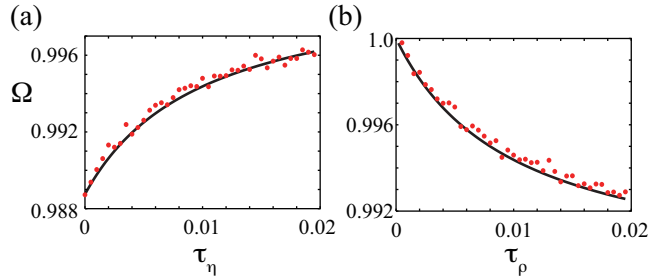}
 \caption{
 Mean frequency $\Omega$ of Stuart-Landau oscillators driven by
 Ornstein-Uhlenbeck processes when $\bm{G}=(x, 0)$, $\sigma=0.3$,
 $\omega=1$ and $c=1$. Solid lines are theoretical predictions,
 Eq. (\ref{omg}). (a) $\tau_\rho=0.01$. (b) $\tau_\eta=0.01$.
 }
 \label{figure:2}
\end{figure}
%%%%%%%%%%%%%%%%%%%%%%%%%%%%%%%%%%%%%%%%%%%%%%%%%%%%%%%%%%%%%%%%%%%%%%

\end{document}